\documentclass[%
 reprint,
 amsmath,amssymb,
aps,
usenames,dvipsnames]{revtex4-1}

\usepackage[version=3]{mhchem} 

\usepackage{dcolumn}
\usepackage{bm}
\usepackage{todonotes}
\usepackage{tikz}
\usepackage{nicefrac}
\usepackage{array}
\usepackage{amssymb}
\usepackage{amsmath}
\usepackage{graphicx}
\usepackage{multirow}
\usepackage{comment}
\usepackage[normalem]{ulem}
\usepackage{url}
\usepackage{mathrsfs}
\usepackage{subcaption}
\usepackage{bbold}
\usepackage{bm}
\usepackage{braket}
\usepackage{soul,xcolor} 
\setstcolor{blue}

\newcommand{\rev}[1]{{\color{black}{#1}}}

\renewcommand{\t}[1]{\textrm{\scriptsize #1}}

\graphicspath{
  {figures/}
}

\begin{document}

\title{Anharmonic Effects in the Low-Frequency Vibrational Modes of Aspirin and Paracetamol Crystals}
\author{Nathaniel Raimbault$^1$}
\author{Vishikh Athavale$^2$}
\author{Mariana Rossi$^1$}
\email{rossi@fhi-berlin.mpg.de\\ nathanielraimbault@gmail.com}
\affiliation{%
 $^1$Fritz-Haber-Institut der Max-Planck-Gesellschaft, Faradayweg 4-6, D-14195 Berlin, Germany \\
 $^2$Department of Chemistry, University of Pennsylvania, Philadelphia, Pennsylvania 19104, United States
}%

\date{January 2019}

\begin{abstract}
The low-frequency range of vibrational spectra is sensitive to collective vibrations of the lattice. In molecular crystals, it can be decisive to identify the structure of different polymorphs, and in addition, it plays an important role on the magnitude of the temperature-dependent component of vibrational free energy differences between these crystals. In this work we study the vibrational Raman spectra and vibrational density of states of different polymorphs of the flexible Aspirin and Paracetamol crystals based on dispersion-corrected density-functional theory, density-functional perturbation theory, and {\it ab initio} molecular dynamics. We examine the effect of quasi-harmonic lattice expansion and compare the results of harmonic theory and the time correlation formalism for vibrational spectra. \rev{Lattice expansion strongly affects the collective vibrations below 300 cm$^{-1}$, but it is significantly less important at higher frequencies, while thermal nuclear motion can be important in the full vibrational range.} We also observe that the inclusion or neglect of many-body van der Waals dispersion interactions do not cause large differences in the low-frequency range of Raman spectra or vibrational density of states, provided the lattice constants are fixed. We obtain quantitative agreement with experimental room-temperature Raman spectra below 300 cm$^{-1}$ for all polymorphs studied. Examining the two-dimensional correlations between different vibrations, we find which modes show a larger degree of anharmonic coupling to others, providing a possible route to assess the accuracy of harmonic free energy evaluations in different cases.
\end{abstract}

\maketitle

\section{Introduction}

\rev{Molecular crystals are a large class of crystals that encompasses} common painkillers and antipyretics like Aspirin, Paracetamol, or Ibuprofen. As suggested by their name, such crystals are built from individual molecular units, which are mainly held together by non-covalent interactions like hydrogen bonds and dispersion forces.
The molecular units that constitute them can be arranged in different patterns \rev{and} each specific arrangement is called a polymorph.
\rev{Despite} the often small energy differences separating these polymorphs~\cite{CCDC_blind_test_2016}, they can present very different physicochemical properties. 
For instance, Paracetamol form II is known to be more soluble than form I, and is also more easily compressible into tablets~\cite{Thomas-Wilson_2011}. Being able to accurately grasp the energetic balance between different polymorphs and to unambiguously characterize them could potentially lead to reduced costs in the pharmaceutical industry, for example. Doing so is no easy task, though: the energy differences between different polymorphs is typically of the order of only a few meV per molecular unit~\cite{CCDC_blind_test_2016} \rev{and vibrational structural fingerprints can show only small (but important) differences}. Therefore, several factors that compete between each other, like anharmonic effects in lattice expansion, nuclear vibration, dispersion forces, and polarization of hydrogen-bonds, \rev{need to be taken into account}~\cite{Hoja_advancedreview_2016,Mariana_PRL_2016,Hoja_FaradayDiscussion_2018}.

\begin{figure}[htb]
  \centering
  \includegraphics[width=0.95\columnwidth]{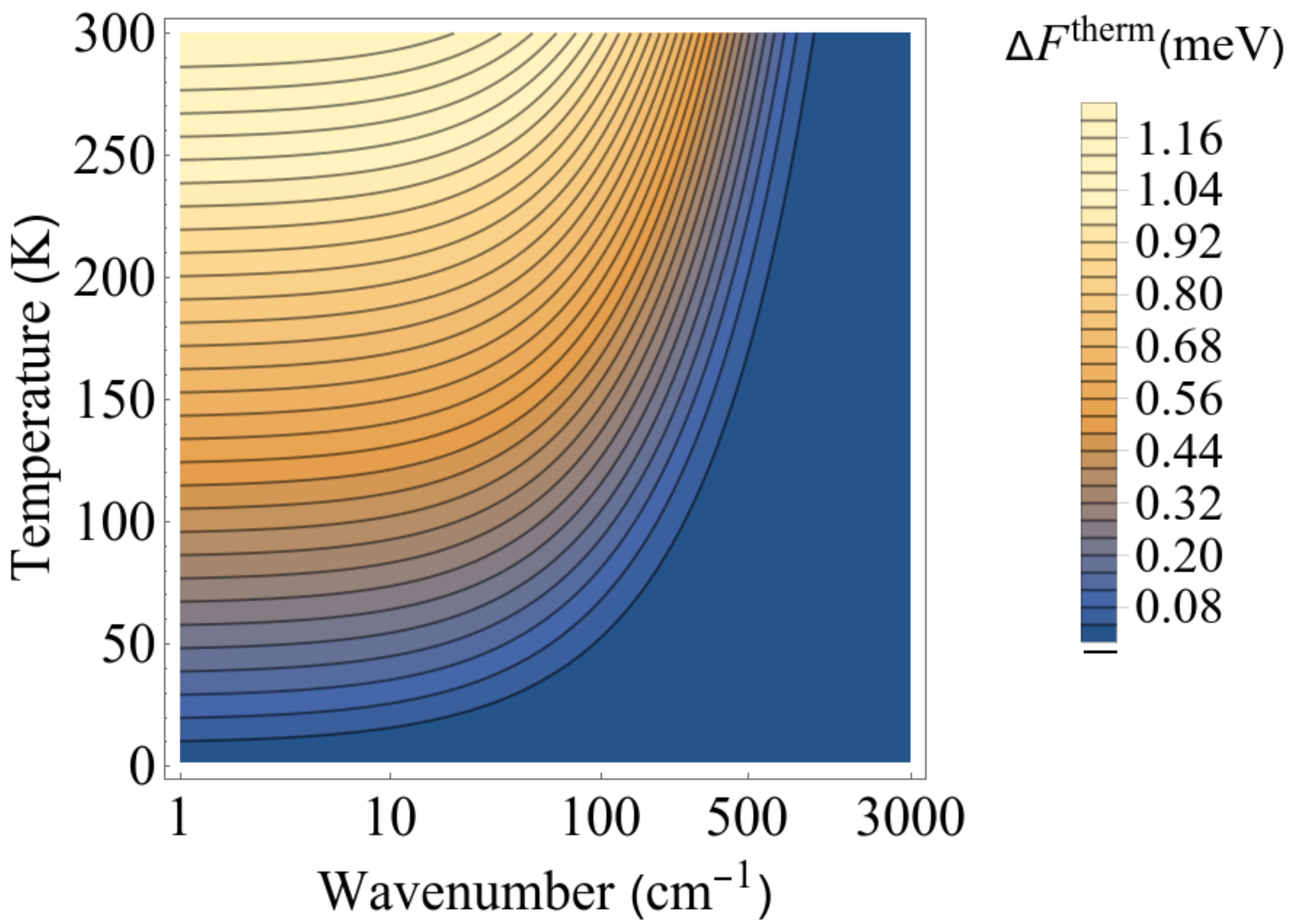}
  \caption{Errors in the temperature-dependent term of the harmonic quantum vibrational free energy \rev{as a function of temperature $T$ and frequency $\omega$, caused by a a percentage error $\Delta \omega$.} \rev{$\Delta F^{\text{therm}}(\omega,T)=k_B T \ln\left[\left(1-e^{-\frac{\hbar(\omega+\Delta \omega)}{k_B T}}\right)/\left(1-e^{-\frac{\hbar\omega}{k_B T}}\right)\right]$. We considered $\Delta \omega = 0.05 \omega$.}}
  \label{fig:fes-error}
\end{figure}

Given the \rev{large} unit-cells \rev{(containing hundreds of atoms)} of some of these molecular crystals, few studies have treated all of these effects in first-principles theoretical calculations~\rev{\cite{Graeme_CrysRev_2011,CCDC_blind_test_2016,Mariana_PRL_2016,Hoja_advancedreview_2016,Price_2009,Price_DDT_2016}}.
Specifically, the impact of anharmonic terms of the potential energy surface (PES) in the temperature-dependent properties of these crystals has only recently started to be addressed~\rev{\cite{Madsen_ACSA_2013,Mariana_PRL_2016,Brela_JPCB_2016,Ruggiero_2017, Hoja:2019,Cervinka_JPCA_2016,Brandenburg_JPCL_2017}}.
In particular, the low-frequency phonon modes, mainly governed by intermolecular interactions (e.g., hydrogen bonds), are sensitive to 
(anisotropic) lattice expansion at finite temperatures~\cite{Hoja_advancedreview_2016,Beran_ACR_2016,Ruggiero_2017} and also to nuclear fluctuations -- \rev{even if the extent of this sensitivity has not yet been carefully quantified}. This region is particularly important since it strongly contributes to the vibrational free energy at finite temperatures~\cite{Rossi_JPCB_2013}. As an illustration, we show in Fig. \ref{fig:fes-error}, the error in the temperature-dependent part of the harmonic vibrational free energy given by a \rev{5\%} error in the vibrational density of states at different frequencies. \rev{Errors in the lower frequencies, especially below 500 cm$^{-1}$} have a large impact in this term, which becomes more pronounced as the temperature increases.

As vibrational spectra can be measured with high accuracy and at different temperatures, comparing theoretical and experimental spectra in the low-frequency region is important to gauge whether temperature-dependent vibrational free energies can be accurately described by any given theoretical methodology.

\begin{figure*}[ht]
  \centering
  \includegraphics[width=2\columnwidth,angle=0]{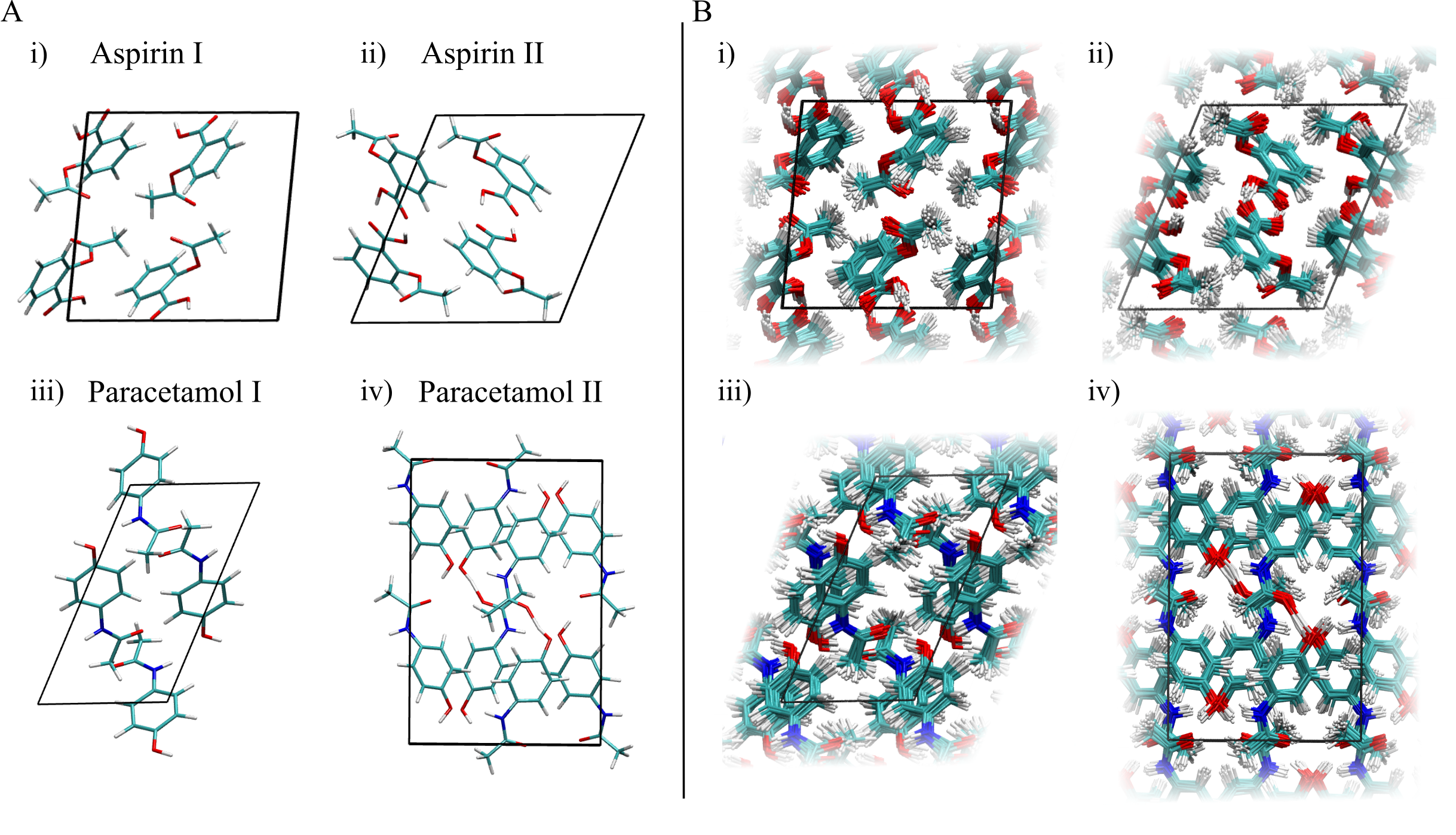}
  \caption{Panel A: The 4 polymorphs of this study at equilibrium: i) Aspirin form I, ii) Aspirin form II, iii) Paracetamol form I, and iv) Paracetamol form II. The unit cell is drawn in black. Panel B: superposition of multiple frames coming from an aiMD simulation. In all systems, methyl groups rotate freely. In Aspirin form II, the distance between neighbouring acetyl groups varies over time.}
  \label{fig_polymorphs_multiframes}
\end{figure*}

In this paper, we present a theoretical characterization of the low-frequency ($\omega <  300$ cm$^{-1}$) vibrational Raman spectra of the two main polymorphs of Paracetamol and Aspirin. Comparing these particular polymorphs is enlightening since while for Paracetamol the hydrogen bonding pattern in the different polymorphs is quite diverse, in Aspirin they are almost identical, as shown in Fig. \ref{fig_polymorphs_multiframes}A. We employ density-functional theory (DFT) and density functional perturbation theory~\rev{\cite{Gonze_APS_1997-1,Gonze_APS_1997-2,Gerratt_JCP_1968,DFPT_dielectric_2018}} (DFPT), including many-body van der Waals corrections\rev{ \cite{Tkatchenko_PRL2012,Ambrosetti_JCP2014}} (MBD). We present an analysis of how low-frequency vibrational modes change with anharmonic couplings to other modes, with changes in the lattice, and with changes in the potential energy surface.

\section{Methods}

In the following, we provide details about the different methodologies we use, as well as the numerical settings we employ in each case.
All our calculations were performed within FHI-aims~\cite{Blum_2009}, an all-electron numeric atom centered orbitals code. We obtained the experimental crystal structures from Ref.~\cite{Wilson_2002} for Aspirin I, Ref.~\cite{Bond_not_James_2011} for Aspirin II, and from Refs. \cite{Drebushchak2004, Wilson2000} for Paracetamol I and II. We compute energies and forces with the PBE exchange-correlation functional throughout, including MBD dispersion corrections as described in Ref. \cite{Distasio_Tkatchenko_2014}, except where stated otherwise. For the Raman spectra, we calculate the polarizability tensors with DFPT \cite{DFPT_dielectric_2018}. We calculate the tensors with the LDA functional, given that we have previously shown in Ref.~\cite{DFPT_dielectric_2018} that it saves considerable computational time and the Raman spectra show no differences when calculating these tensors with different functionals. In the following, it will thus always be assumed that LDA was used for calculating polarizabilities, even if not explicitly mentioned. Unless stated otherwise, a $2\times2\times2$ $k$-point grid was used for all polymorphs. 

The data presented in this work as well as the input and output files used to produce it are publicly available as a dataset~\cite{nomad} in the NOMAD Repository.

\subsection{Lattice Expansion \label{sec:expansion}}

In order to assess anisotropy in the quasi-harmonic lattice expansion calculations we assumed fixed angles for each molecular crystal polymorph and minimized the second order Taylor-expansion of the Helmholtz free energy $F$ at a particular temperature $T$,
\begin{align}
    F(a,b,c)=F_0 + \mathbf{p}^t \, \mathbf{H} \,  \mathbf{p} \,, \label{eq:quasihm}
\end{align}
\noindent where $F_0$ is the free energy at the equilibrium lattice parameters at the temperature of choice, $\mathbf{H}$ is the matrix of second derivatives of the free energy with respect to the lattice parameters, $\mathbf{p}=(a-a_0, b-b_0, c-c_0)$ where $a_0$, $b_0$ and $c_0$ are the equilibrium unit-cell parameters at a given temperature, and $\mathbf{p}^t$ its transpose. \rev{Given the symmetric nature of $\mathbf{H}$, we thus have 10 unknown variables to determine at a specific temperature $T$, namely the 6 $\mathbf{H}$ matrix elements, the 3 equilibrium lattice parameters, and $F_0$. This minimization was done at the assumption of fixed cell angles and employing a non-linear least-square fit. We used at least 10 evaluations of $F$ calculated at different values of $a$, $b$ and $c$ in the harmonic approximation from DFT (at any given temperature) in order to determine these parameters. Once DFT phonons are obtained for different lattice displacements the temperature dependence of the vibrational free energy in the harmonic approximation is analytically known, thus making it easy to find the optimum parameters of Eq. \ref{eq:quasihm} at any temperature}. Further details of this procedure can be found in the SI.
We note that a possible alternative, which would require a comparable amount of phonon evaluations, would be to evaluate mode-specific Gr{\"u}neisen parameters~\cite{Heit_Beran_ACSB_2016} to approximate \rev{the} variation of phonon frequency with each lattice parameter.

The \textit{tight} basis sets of FHI-aims were used for all atomic species and the phonon calculations using a $1 \times 1 \times 1$ supercell and a $5 \times 5 \times 5$ $q$-point grid were performed using the Phonopy program~\cite{phonopy}. We calculated 14 distortions of the lattice for Aspirin I, 10 for Aspirin II, 15 for Paracetamol I and 13 for Paracetamol II.  \rev{We checked that the solution is a minimum of the free-energy surface by ensuring that the eigenvalues of $\mathbf{H}$ are all positive. Whenever that was not the case, we displaced the cell in the direction of the eigenvector with the negative eigenvalue, calculated phonons for this displacement and added this point to the minimization procedure. The routine written for this model is available online~\cite{github}.} 

\subsection{Harmonic Raman Spectra}
\label{sec:ha_ram}

A standard way to evaluate vibrational Raman spectra is through the harmonic approximation, in which the Taylor expansion of the potential energy is truncated at the second order; In this procedure, Raman intensities are proportional to the derivatives of the polarizability tensor with respect to atomic displacements (see e.g., Refs.~\cite{Neugebauer_JCC2002,Veithen:2005gf}).
We calculate the orientation-averaged ``powder" harmonic Raman intensity $I_H(\omega)$ (which is directly proportional to the Raman scattering cross section~\cite{Murphy1989}) of a given normal mode $p$ by~\cite{Prosandeev2005}, 

\begin{align}
I_{H}(\omega) & = I^{\perp}_H+I^{\parallel}_H \propto \frac{1}{\omega(1-e^{-\beta\hbar\omega})} \frac{1}{30} (10 G_p^{(0)}+7 G_p^{(2)}) \,, \label{eq:harm-raman} \\
\nonumber G_p^{(0)} & =\frac{1}{3}\left[ ({\alpha}_{xx}')_p + ({\alpha}_{yy}')_p + ({\alpha}_{zz}')_p \right]^2 \\
\nonumber G_p^{(2)} & =\frac{1}{2} \left[(2 {\alpha}_{xy}')_p^2 + (2 {\alpha}_{xz}')_p^2 + (2 {\alpha}_{yz}')_p^2\right] + \\ \nonumber & + \frac{1}{3} \{ \left[({\alpha}_{xx}')_p-({\alpha}_{yy}')_p\right]^2 +  \\ \nonumber & + \left[({\alpha}_{xx}')_p-({\alpha}_{zz}')_p\right]^2 + \left[({\alpha}_{yy}')_p-({\alpha}_{zz}')_p\right]^2 \}
\end{align}
where $I^{\perp}$ and $I^{\parallel}$ are the depolarized and polarized Raman intensities respectively, $\beta = 1/(k_B T)$, $({\alpha}_{ij}')_p=\left(\partial {\alpha}_{ij}/\partial Q_p\right)_0$ is the derivative of the $ij$ component of the polarizability with respect to the displacement of normal mode $Q_p$.
We compute these derivatives through finite differences, in which we evaluate the polarizability tensor from DFPT~\cite{DFPT_dielectric_2018} at $6N$ (forward and backward) nuclear displacements in the unit cell around the equilibrium position, $N$ being the number of atoms per unit cell.
\rev{We use regular cartesian coordinates to describe normal modes. This coordinate system may not always be appropriate when dealing with torsional vibrational modes, for which a better approach is to use curvilinear coordinates}~\cite{Rybkin_2014}.
Knowing the space group of our crystals, we then apply all symmetry operations pertaining to this group onto the vibrational modes, in order to confirm whether they \rev{should} show Raman activity or not~\footnote{From group theory, for a mode to be Raman active, it needs to have the same symmetry as a component of the polarizability tensor, i.e. $x^2$, $y^2$, $z^2$, $xy$, $xz$ or $yz$.}. \textit{A posteriori}, we discard modes that are Raman-inactive, if any are found.
\rev{We note, however, that selection rules are naturally enforced by the whole procedure, and such an \textit{a posteriori} correction is only needed if (rare) numerical errors are present.}

Harmonic Raman spectra presented in this paper were calculated with \textit{light} numerical and basis-set settings in the FHI-aims code~\cite{Blum_2009}, for direct comparison with the anharmonic spectra. Differences between {\it light} and {\it tight} harmonic spectra are minor and shown in \rev{Fig. S3} the SI.

\subsection{Anharmonic Raman Spectra}
\label{sec:anh_ram}
A way to go beyond the harmonic approximation is to resort to the time correlation formalism. \rev{In this formalism, the potential energy is sampled without resorting to any approximations, thus ensuring that full anharmonicity (here at a classical nuclei approximation) is captured}. Vibrational Raman lineshapes are proportional to the Fourier transform of the static polarizability (or static dielectric) autocorrelation function~\cite{Mukamel_1995}. Realizing this formalism requires computing polarizability tensors along molecular dynamics trajectories, as explained in, \rev{e.g.,} Ref.~\cite{DFPT_dielectric_2018}.
For comparison with experimental spectra taken from powder samples, we here calculate the anharmonic powder averaged Raman intensity $I_A(\omega)$. This can be calculated from the isotropic and anisotropic contributions to the signal as follows (see Ref.~\cite{McQuarrie_2000_statistical}, Eqs. 21-99 and 21-100, and also~\cite{Murphy1989}), 

\begin{align} \label{eq_raman_intensity}
I_A(\omega) & =  I^{\perp}_A + I^{\parallel}_A = I_\t{iso}(\omega) +\frac73 I_\t{aniso} (\omega) \\
\nonumber I_\t{iso}(\omega) & = I^{\parallel}_A - \frac43 I^{\perp}_A \propto  \int_{-\infty}^{+\infty}dt e^{-i\omega t} \langle\bar{\alpha}(0)\bar{\alpha}(t)\rangle  \\
\nonumber I_\t{aniso} (\omega) & = I^{\perp}_A \propto    \int_{-\infty}^{+\infty}dt e^{-i\omega t} \frac{1}{10}\langle \textrm{Tr}[\bm{\beta}(0)\cdot \bm{\beta}(t)]\rangle \,,
 \end{align}

\noindent where the polarizability (or dielectric) tensor $\bm{\alpha} = \bar{\alpha} \bm{I} + \bm{\beta}$ and the brackets $\langle.\rangle$ denote the canonical average. \rev{We consider here only the electronic contribution to the dielectric permittivity. In ionic and polar crystals, contributions from ionic polarization could be more pronounced, as discussed in Ref.~\cite{Calzolari2013}.} 
We apply the so-called quantum or Kubo-transform correction factor to the lineshapes of $\beta \hbar \omega/(1-e^{-\beta \hbar \omega})$.

All our {\it ab initio} molecular dynamics (aiMD) simulations were performed with \textit{light} numerical and basis sets settings in FHI-aims and otherwise the same settings as for all other calculations. We performed a thermalization (NVT) run of about 2 picoseconds for each polymorph, followed by two NVE simulations of 15 picoseconds each, using a time step of 0.5 femtosecond. We computed polarizability tensors with DFPT calculations every 1 fs.

\subsection{2D-correlation spectra \label{sec:2d}}

Two-dimensional (2D) correlation spectra can be calculated from our MD simulations. We follow the procedure detailed in Ref.~\cite{Noda_2000}. These 2D correlations ``provide a quantitative comparison of spectral density variations observed at two different spectral variables over some finite observation interval $T_{min}$ and $T_{max}$"~\cite{Noda_2000}. It is thus a useful tool to understand couplings between different modes, as has been recently shown for the case of water~\cite{Morawietz_2018}. We recall the main formulae from Ref.~\cite{Noda_2000} to produce such spectra in the present text.
We define the dynamic spectrum $\tilde{I}_j(\omega)$ as
\begin{align}
    \tilde{I}_j(\omega)=I_j(\omega)-\bar{I}(\omega) \,,
\end{align}
where $I_j(\omega)$ is simply the intensity (e.g., Raman intensity as defined in Eq.~\ref{eq_raman_intensity}) of at $\omega$, evaluated for a given time window $T_{win}$, the duration of which depends on the phenomenon one wants to observe.
If we divide our trajectory of interval $[T_{min},T_{max}]$ into $m$ segments of length $T_{win}$ evenly spaced by an increment of $(T_{max}-T_{min})/(m-1)$, the average spectrum $\bar{I}(\omega)$ is given by 
\begin{align}
    \bar{I}(\omega)=\frac{1}{m} \sum_{j=1}^m I_j(\omega)\,,
\end{align}
and the synchronous 2D correlation intensity takes the following expression,
\begin{align}
    \phi(\omega_1,\omega_2)=\frac{1}{m-1}\sum_{j=1}^m \tilde{I}_j(\omega_1)\tilde{I}_j(\omega_2) \,.
\end{align}
In all our simulations, we choose $\rev{T_{\text{win}}}=1$ ps. 

The diagonal peaks appearing in these spectra are referred to as autopeaks and are always positive; They represent the change in intensity at a given frequency over a given period of time. Hence, regions that change intensity a lot will have strong autopeaks, while regions that vary little will have weak autopeaks.
Off-diagonal peaks, or cross-peaks, correspond to simultaneous changes (of equal or opposite signs) in intensities at two different frequencies over a given duration, indicating a possible coupling between the two corresponding vibrational modes.

\section{Results}
\subsection{Lattice Expansion and Harmonic Raman Spectra}

The four molecular crystals studied in this work, namely Aspirin I and II, and Paracetamol I and II, have known crystal structures which are shown in Fig.~\ref{fig_polymorphs_multiframes}A.
In many molecular-crystal polymorphs, important structural differences can already be spotted simply by looking at the molecular arrangement and the shape of the unit cell\rev{, even without resorting to Raman spectroscopy.
Paracetamol I and II are a good example of such different polymorphs, as shown in Figure \ref{fig_polymorphs_multiframes}A.}
For a few, however, differences are much more subtle. This is particularly true for Aspirin, for which both forms appear to have the same structure in projection. One key difference lies in the pattern formed by intermolecular hydrogen bondings between the acetyl groups (CH$_3$CO)~\cite{Bond_not_James_2011,CROWELL201529,Brog2013,Vishweshwar2005,Takahashi2014}.

Lattice expansion can have a large impact on the energetics, affecting the stability ranking of polymorphs~\cite{Nyman_2016}. However, we will here focus on its impact on vibrational spectra, and more specifically on Raman spectra. To this end, we  calculate the harmonic Raman spectra (see section~\ref{sec:ha_ram}) of Aspirin I and II, using for each of them two different lattice parameters: the experimentally determined ones obtained from Refs.~\cite{Wilson_2002,Bond_not_James_2011,Drebushchak2004, Wilson2000}, and the ones coming from the procedure outlined in Section \ref{sec:expansion}. The lattice parameters we use are given in Table~\ref{tab_lattice_param}.

  \begin{table*}[htbp]
  \begin{tabular}{l|cccc|cccc|cccc|cccc|ccc}\hline\hline
  & \multicolumn{4}{c|}{PES (PBE)} & \multicolumn{4}{c|}{PES (PBE+MBD)}  & \multicolumn{4}{c|}{Calc. 300K} & \multicolumn{4}{c|}{Exp. 300K} & \multicolumn{3}{c}{$\Delta$ (Calc-Exp) (\%)}\\
  System/parameter & $a$& $b$& $c$& $\beta$ & $a$ & $b$ & $c$ & $\beta$ & $a$&$b$&$c$&$\beta$ &$a$&$b$&$c$&$\beta$  &$\Delta a$&~~$\Delta b$~~&$\Delta c$ \\ \hline
  Paracetamol I &6.92 &12.51 & 12.98&55.8 & 7.01 & 9.15 & 12.77 & 66 & 7.09 & 9.23 & 12.75 & 66 & 7.08 & 9.34 & 12.85 & 64.5 &0.14 &~1.18&0.78  \\
  Paracetamol II &11.63 &9.14 &17.40 & 90& 11.59 & 7.30 & 17.26 & 90  & 11.62 & 7.61 & 17.26 & 90 & 11.83 & 7.40 & 17.16 & 90 &1.78&~2.84&0.58 \\ 
  Aspirin I & 12.30 & 7.00 & 12.30 & 96 & 11.40 & 6.52 & 11.33 & 96  & 11.50 & 6.51 & 11.46 & 96 & 11.42 & 6.60 & 11.48 & 96 &0.70&~1.36&0.17\\ 
  Aspirin II & 13.26&6.85 &12.38 &114 & 12.27 & 6.43 & 11.35 & 111 & 12.52 & 6.65 & 11.82 & 111 & 12.36 & 6.53 & 11.50 & 112 &1.29&~1.84&2.78\\ 
  \hline\hline
 \end{tabular}
 \caption{Lattice parameters. Unit vectors are in \AA, and angles in degrees. Second and third columns: fully (atomic positions and unit cell) optimized structure at the potential energy surface (PES) using the PBE and the PBE+MBD functional. Fourth column: calculated lattice constants from our quasiharmonic lattice expansion scheme, \rev{calculated with the PBE+MBD functional}. Fifth column: experimental lattice constants at 300K from Refs.~\cite{Wilson_2002,Bond_not_James_2011,Drebushchak2004, Wilson2000}. Last column: error (in percentage) between the calculated and experimental lattice parameters at 300K.}
 \label{tab_lattice_param}
 \end{table*}
We observe that most calculated lattice parameters at 300 K are relatively close to the experimental 300 K results, although the changes are quite heterogeneous. The most notable differences can be seen for the $b$ parameter of Paracetamol II and the $c$ parameter of Aspirin II, that show both a difference of about 2.8\%. The absolute values of the calculated lattice constants at 300 K for form I of Paracetamol and Aspirin are closer to experiment than those of their respective form II, although the relative expansion is better reproduced for the latter (see SI, \rev{Fig.~S1}).
Other works have been conducted on a similar topic, but employing different functionals and a different approach, as they fixed the experimentally-determined volumes at different temperatures and only optimized the lattice parameters, leading to an expected good agreement with experimental values~\cite{ADHIKARI2015109}. \rev{We draw attention to the fact that lattice parameters calculated without van der Waals dispersion interactions (at the potential energy surface), shown in Table \ref{tab_lattice_param}, strongly deviate from experimental values, highlighting the importance of these interactions in determining the shape and density of these crystals.}

\begin{figure}[htbp]
  \centering
  \includegraphics[width=1.0\columnwidth]{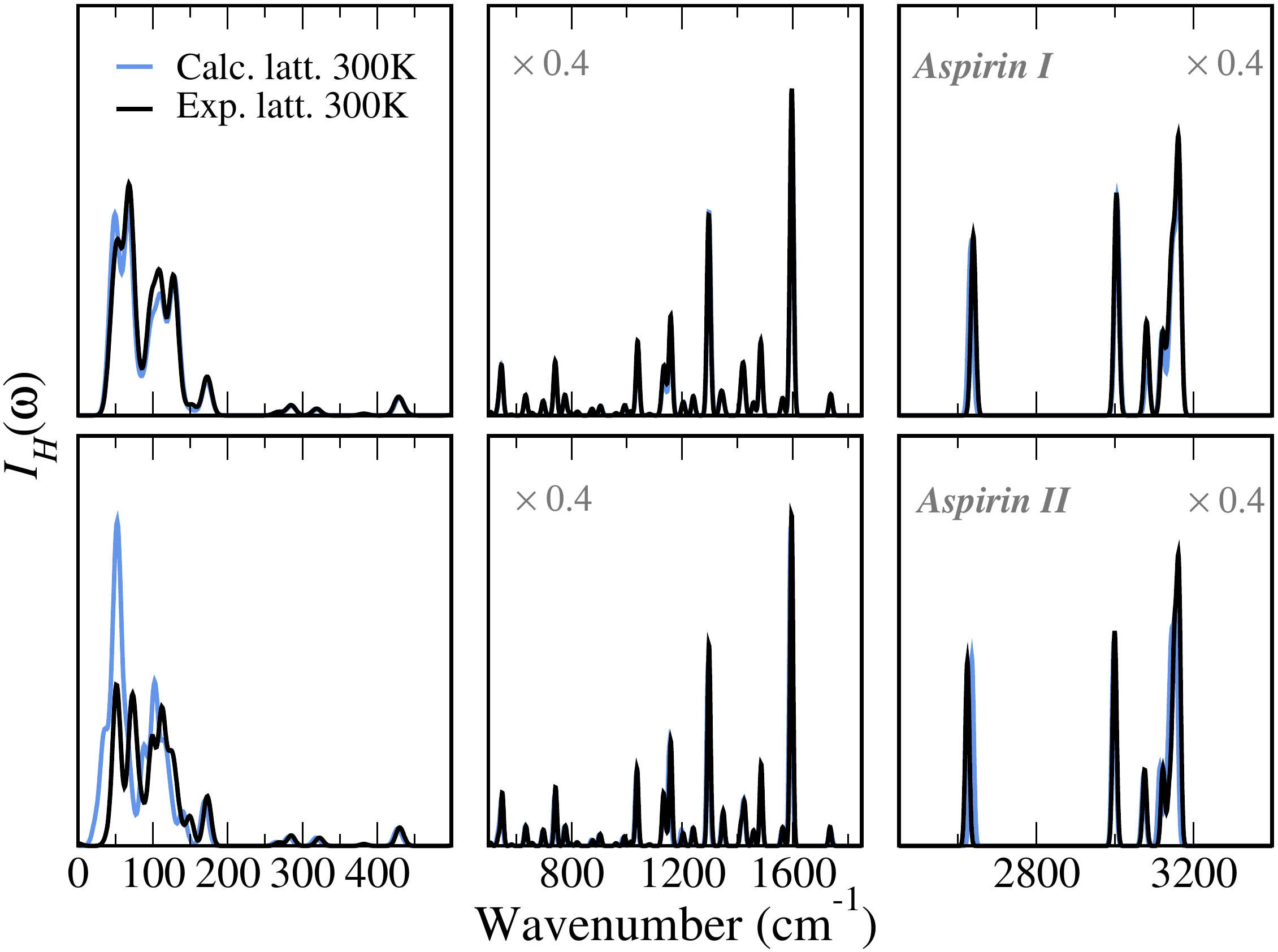}
  \caption{Harmonic Raman spectra of Aspirin I (top) and II (bottom) obtained with the experimental lattice parameters (black) and the calculated lattice parameters (blue) at 300 K. We used the PBE+MBD functional. In the middle and right panels, we indicate the factor by which each panel is scaled as compared to the left panel.}
  \label{fig_asp_ha_lattice}
\end{figure}

The harmonic Raman spectra of Aspirin form I and II computed with our calculated lattice parameters \rev{and the experimental ones} at 300 K are shown in  Fig.~\ref{fig_asp_ha_lattice}.
\rev{We observe that using different lattice parameters has seemingly no influence in the middle-frequency range, while at high frequencies, only small rigid shifts and minor changes in intensities can be observed.}
The impact is most easily seen at low frequency, especially for form II, where the intensities of several peaks are modified, and also the general lineshape changes substantially.

It is not surprising to see that the changes manifest mostly at low frequency, since, as mentioned in the introduction, low-frequency phonon modes tend to be sensitive to the intermolecular potential, which is determined by the shape of the lattice.
The differences between the two polymorphs can stem from different factors. As we have previously seen, the difference between the experimental lattice constants and the calculated ones is greater for Aspirin II than Aspirin I, so it is only logical that this difference is reflected in the spectrum.
Also, it is known that Aspirin II is more easily compressible, as it forms flat hydrogen-bonded sheets along the $c$ axis, as opposed to a wave-like pattern in form I~\cite{Brog2013}. This seems to be consistent with the larger expansion of the $c$ lattice vector of Aspirin II, which is observed both in our calculations and in experiments, even though the differences between both polymorphs are rather small.
In any case, it is very interesting to notice that relatively \rev{moderate} changes in lattice parameters can translate to a \rev{more} noticeable change on the low-frequency range of the harmonic Raman spectrum.

\begin{figure}[htbp]
  \centering
  \includegraphics[width=1.0\columnwidth]{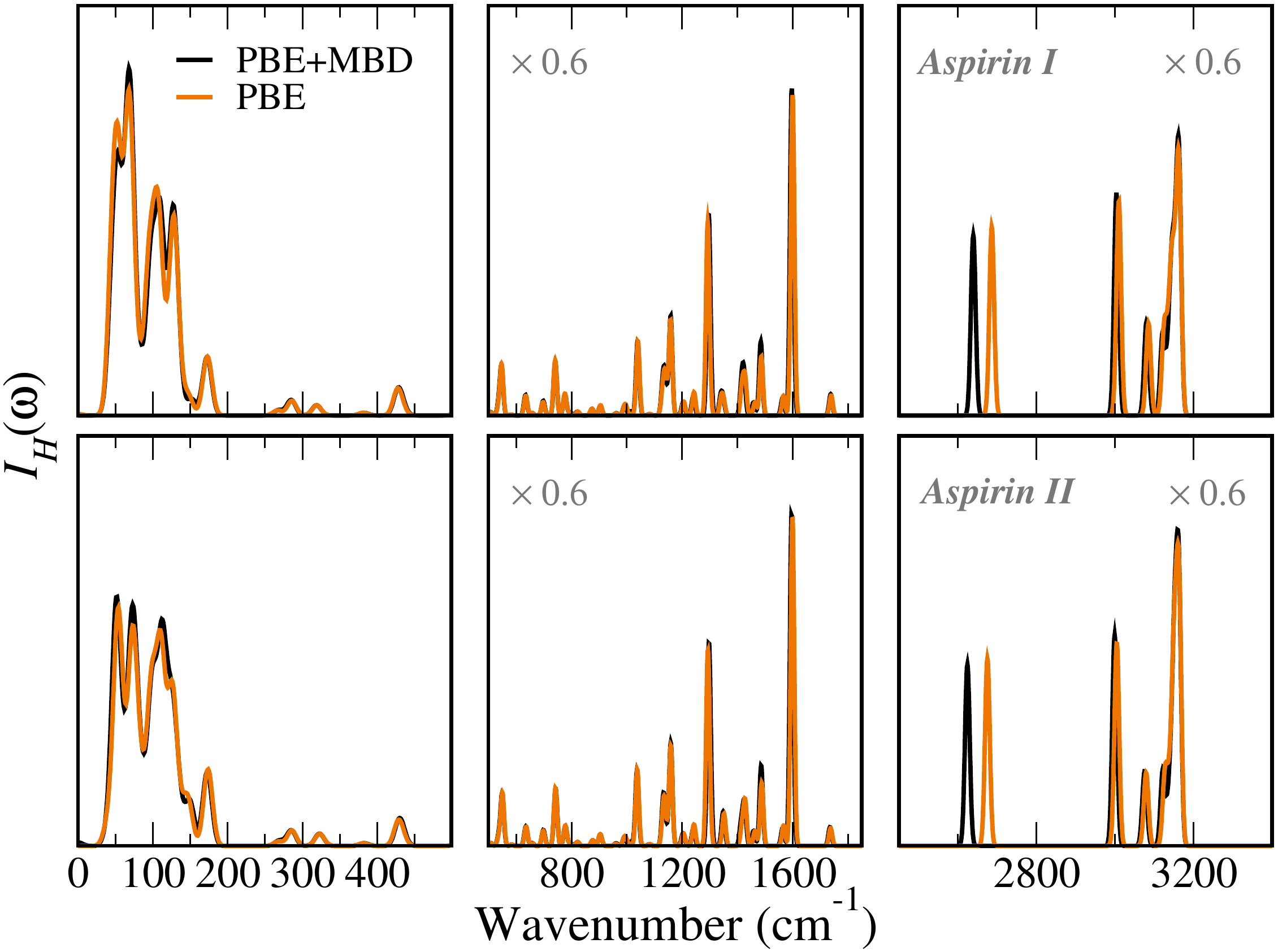}
  \caption{Harmonic Raman spectra of Aspirin I (top) and II (bottom) obtained with the PBE (orange) and the PBE+MBD (black) functionals, using the 300 K experimental lattice constants.}
  \label{fig_asp_ha_mbdornot}
\end{figure}

The lattice is not the only parameter that may impact a vibrational spectrum.
Another worthwhile aspect to consider is the exchange-correlation functional that is used\rev{, or the addition of vdW dispersion}. Especially for systems such as molecular crystals, dispersion forces are known to play an important role and change the energetics substantially~\cite{Hoja_FaradayDiscussion_2018}. We report the impact of dispersion interaction on the harmonic Raman spectra of the same Aspirin polymorphs in Fig.~\ref{fig_asp_ha_mbdornot}. In order to decouple different effects, we maintain the experimental lattice parameters in this case, but fully optimize the atomic positions 
with the different \rev{potential energy surfaces}.

We observe that at high frequencies, ignoring van der Waals contributions in this case results in a blueshift of 47 (51) cm$^{-1}$ of the peak located at 2639 (2624) cm$^{-1}$ for Aspirin I (II). This peak corresponds to symmetric O-H stretching \rev{motions} between the molecular dimers.
We note that we observe a similar shift of this band when simulating these spectra using 300 K and 123 K experimental lattice parameters (see SI, \rev{Fig. S2}). These observations are consistent with the fact that removing van der Waals interactions from the model weakens the H-bonds, and so does increasing the temperature. Consequently, these two effects lead to a blue-shift of this band.
\rev{The middle- and low-frequency ranges remain basically unaltered by the change of functional.
We note that in} the geometries considered here, we observe the presence of a vibrational mode at 33 cm$^{-1}$ in both polymorphs when including MBD corrections. The same modes are at 34 cm$^{-1}$ when neglecting MBD corrections.
These are not Raman-active modes, though, and hence do not show up in Fig.~\ref{fig_asp_ha_mbdornot}.

The results presented so far confirm the importance of taking lattice expansion into account when assessing Raman spectra. The discrepancies we observe in our calculated lattice constants at 300 K, in comparison to experiment, can be due to several factors, among them the exchange-correlation functional and the approximations in the lattice-expansion procedure itself (for instance the assumption of fixed angles or the fact that the multi-parameter optimization of Eq. \ref{eq:quasihm} can lead to meta-stable minima). A detailed investigation of this issue requires studying a broader set of crystals, functionals, and a careful benchmark between different methods (including carrying out computationally costly constant pressure aiMD simulations at different temperatures directly). This will be the subject of a future study. For the remainder of the present work, and in order to focus solely on the impact of vibrational anharmonic contributions, we maintain the experimental lattice parameters (see Table \ref{tab_lattice_param}) for Aspirin I, II, Paracetamol I and the one reported in Ref. \cite{Neumann2009} for Paracetamol II.

\subsection{Anharmonic effects on vibrational modes and Raman lineshapes}

\begin{figure}[h]
  \centering
  \includegraphics[width=1.0\columnwidth]{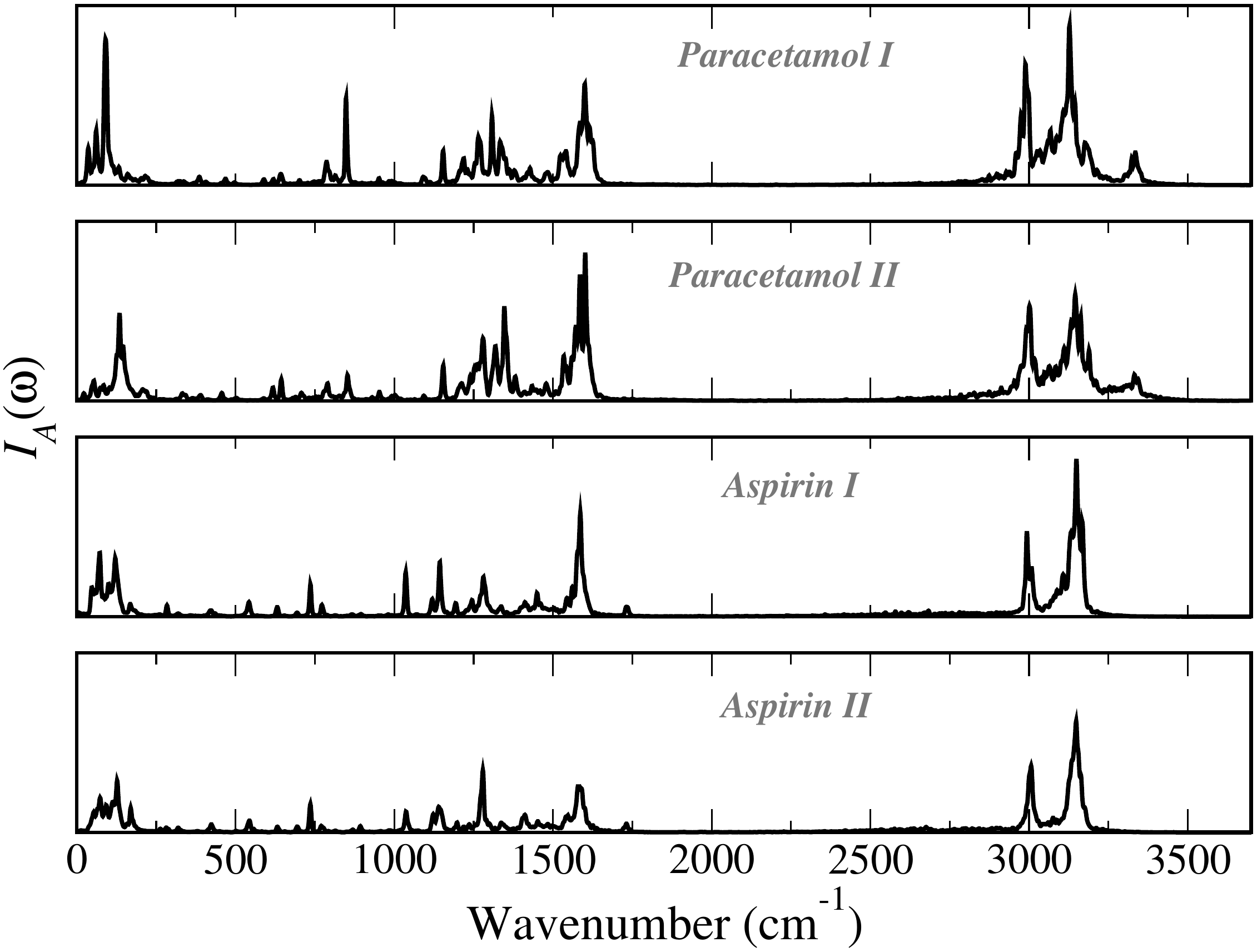}
  \caption{Anharmonic Raman spectra of the four polymorphs of this study on the whole frequency range calculated from aiMD at 300 K with the PBE+MBD functional. Here the intensities have been scaled by a frequency-dependent factor of $\sqrt{\omega}$ solely for visualization purposes\rev{, i.e., so that high frequencies become as visible as low frequencies}.}
  \label{fig_Raman_wholerange}
\end{figure}

As shown in Fig. \ref{fig_polymorphs_multiframes}(b), the pronounced nuclear fluctuations that are observed at room temperature. 
We note, for example, that during aiMD simulations, in all cases the methyl (CH$_3$) groups rotate freely and we observe hydrogen-transfer events between two Aspirin monomers. Other torsional motions within the molecular units are also activated, for example the rotation of the aromatic ring with respect to the rest of the molecules. The question is how these fluctuations translate to the vibrational Raman spectra.

The time correlation formalism gives access to the full anharmonicity of the potential energy surface within the approximation used for the dynamics of the nuclei (e.g., classical or quantum). It is thus able to capture combination bands, overtones and the phonon lifetimes that give rise to the anharmonic lineshape. A drawback of this formalism is that the assignment of vibrational modes is not straightforward and it is often based on the corresponding harmonic spectrum, for which modes are well defined~\cite{Chen_Bowman_PCCP_2018}. Techniques have been proposed to extract effective vibrational modes directly from aiMD simulations~\cite{Mathias_Baer_2011,Mathias_2012}. Such techniques, although very successful in small molecules, require large sampling time\rev{s} and are not straightforward to apply to larger and very flexible systems or in simulations that incorporate nuclear quantum statistics.
\begin{figure}[h]
 \centering
 \includegraphics[width=1.0\columnwidth]{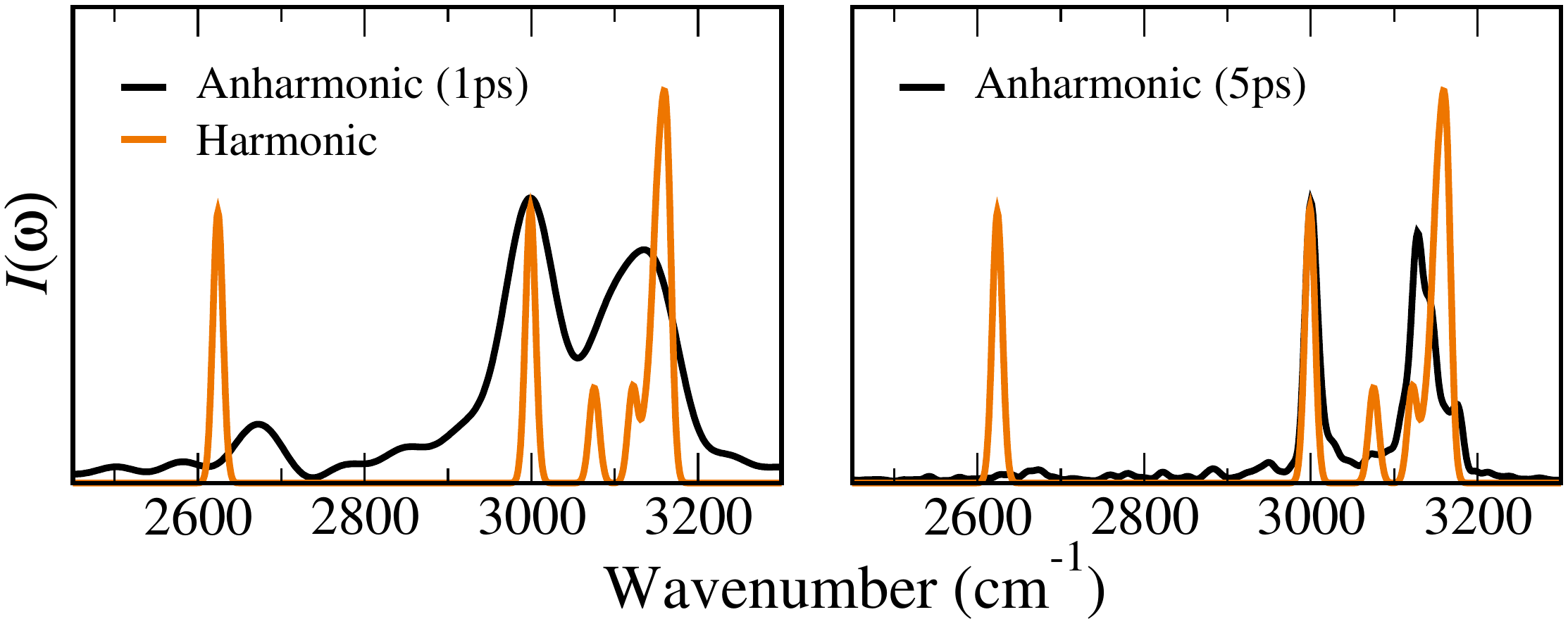}
 \caption{Evolution of the anharmonic Raman spectrum of Aspirin II for different simulation lengths as compared to the harmonic Raman spectrum. In each case, the height of the anharmonic peak located at 3000 cm$^{-1}$ has been adjusted to the harmonic one. Note that the same effect is also observed for Aspirin I.}
 \label{fig_asp_broadpeak}
\end{figure}

\begin{figure*}[ht]
  \centering
  \includegraphics[width=2.0\columnwidth]{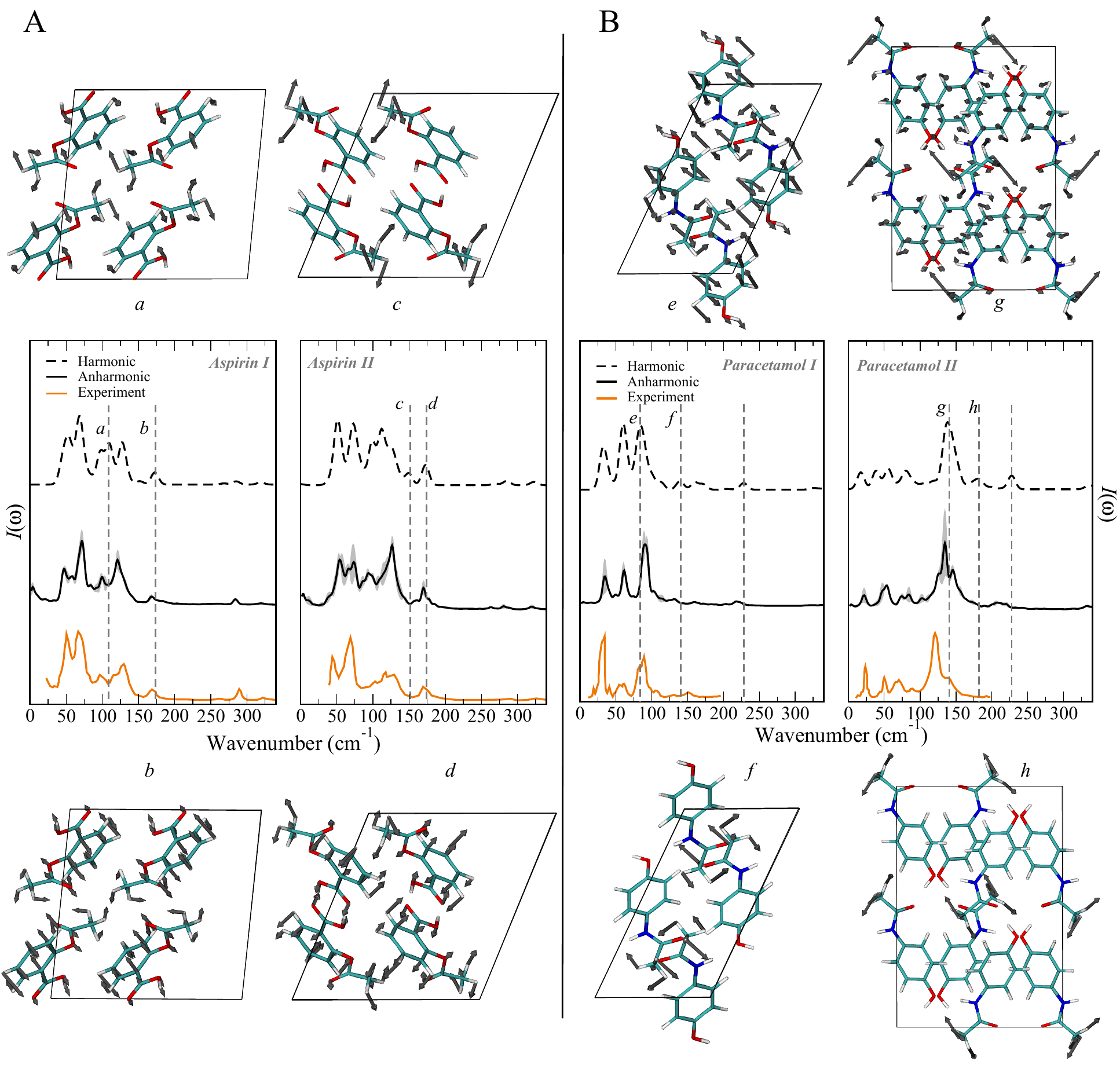}
  \caption{Comparison of the PBE+MBD harmonic, anharmonic (300 K), and experimental (300 K) Raman spectra of Aspirin I and II, and Paracetamol I and II.  The experimental data was extracted from Ref.~\cite{CROWELL201529} and \cite{Roy_2013}. The shaded areas around the anharmonic spectra indicate the uncertainty derived from the statistical error of the different trajectories. Selected normal modes from the harmonic analysis are also shown, for which the arrows represent the direction and the amplitude of the moving atoms. The unit cell is drawn in black.}
  \label{fig_lowrange_poly}
\end{figure*}

In Fig.~\ref{fig_Raman_wholerange}, we show the calculated anharmonic Raman spectra in the full frequency range \footnote{Small differences in the spectra of Paracetamol forms I and II (high-frequency region) with respect to what was published by us in Ref. \cite{DFPT_dielectric_2018} stem from the fact that the lattice constants reported in Ref. \cite{Neumann2009} were used  in that reference, whereas here we use the experimental ones reported in Table \ref{tab_lattice_param}.}. 
The most interesting observation regarding the high-frequency region is that the intense peak observed at 2639 (2624) cm$^{-1}$ for Aspirin I (II) in the harmonic approximation, corresponding to the stretch of O-H bonds that connect the H-bonded Aspirin dimers in the crystal, seems to be completely absent from the anharmonic spectra, and in fact also from experiments as shown in Ref.~\cite{BOCZAR200363}. However, when calculating Raman spectra from short-time (1 ps) autocorrelation functions of the polarizability tensors, one can see that this peak is present, but gets broadened and loses intensity upon increasing simulation time, as shown in Fig.~\ref{fig_asp_broadpeak} (we also show in the SI, Fig. S6, that this vibration is actually present in the vibrational density of states -VDoS- of the hydrogen involved in this mode). 
In addition, this mode is connected to the observed hydrogen transfer events between two Aspirin monomers, which we expect to become more pronounced or turn into a fully shared hydrogen if nuclear quantum effects are included.

As shown in the SI, Fig. S4, and further discussed in the next section, neglecting vdW contributions in the anharmonic Raman spectra of all crystals results in only small changes to the lineshapes and intensities, as long as the lattice parameters are kept fixed at the same values.

In Fig.~\ref{fig_lowrange_poly} we focus on the structure-sensitive low-frequency range of these spectra. We compare our harmonic and anharmonic spectra to experimental results for Aspirin and Paracetamol obtained from powder samples at room temperature, as reported in Refs.~\cite{CROWELL201529} and \cite{Roy_2013}, respectively. For visual comparison, we normalized to 1 the highest peak of each spectrum. \rev{For both systems, there is a good visual agreement between the harmonic Raman spectra and the experimental spectra. There are significant shifts between the harmonic and anharmonic Raman spectra for only a subset of the peaks in this region, which shift closer to the position observed in experiment when anharmonicity is included. 
For clarity we label some of these peaks and depict their harmonic normal mode of vibration in Fig.~\ref{fig_lowrange_poly}, noting that for each system, there are both localized vibrational modes (here typically methyl group rotations) and a more global-motion modes. We will see in the next section that the more pronounced shift of the collective modes correspond to vibrations that have a stronger correlation with other modes.
While none of the approaches allows us to reach a perfect quantitative agreement with experiment, the anharmonic spectra provide an overall better description of peak positions, in particular for Paracetamol I and II. Some of the discrepancies in relative intensities could be resolved by increasing statistical sampling. We highlighted some of the main frequency shifts between harmonic and anharmonic results by placing vertical dotted line.}

\subsection{Mode coupling}

In order to further analyze the effects of anharmonic mode coupling in the low-frequency vibrational range of these crystals, we turn our attention to the vibrational density of states (VDoS), which we here also calculate within the time-correlation formalism, by summing the Fourier transforms of the atomic velocity autocorrelation functions. This quantity plays a more direct role in the estimation of vibrational free energies of these crystals. Experimentally, only inelastic neutron scattering can directly access the VDoS and such measurements are rare for most materials.
\begin{figure}[h]
  \centering
  \includegraphics[width=1.0\columnwidth]{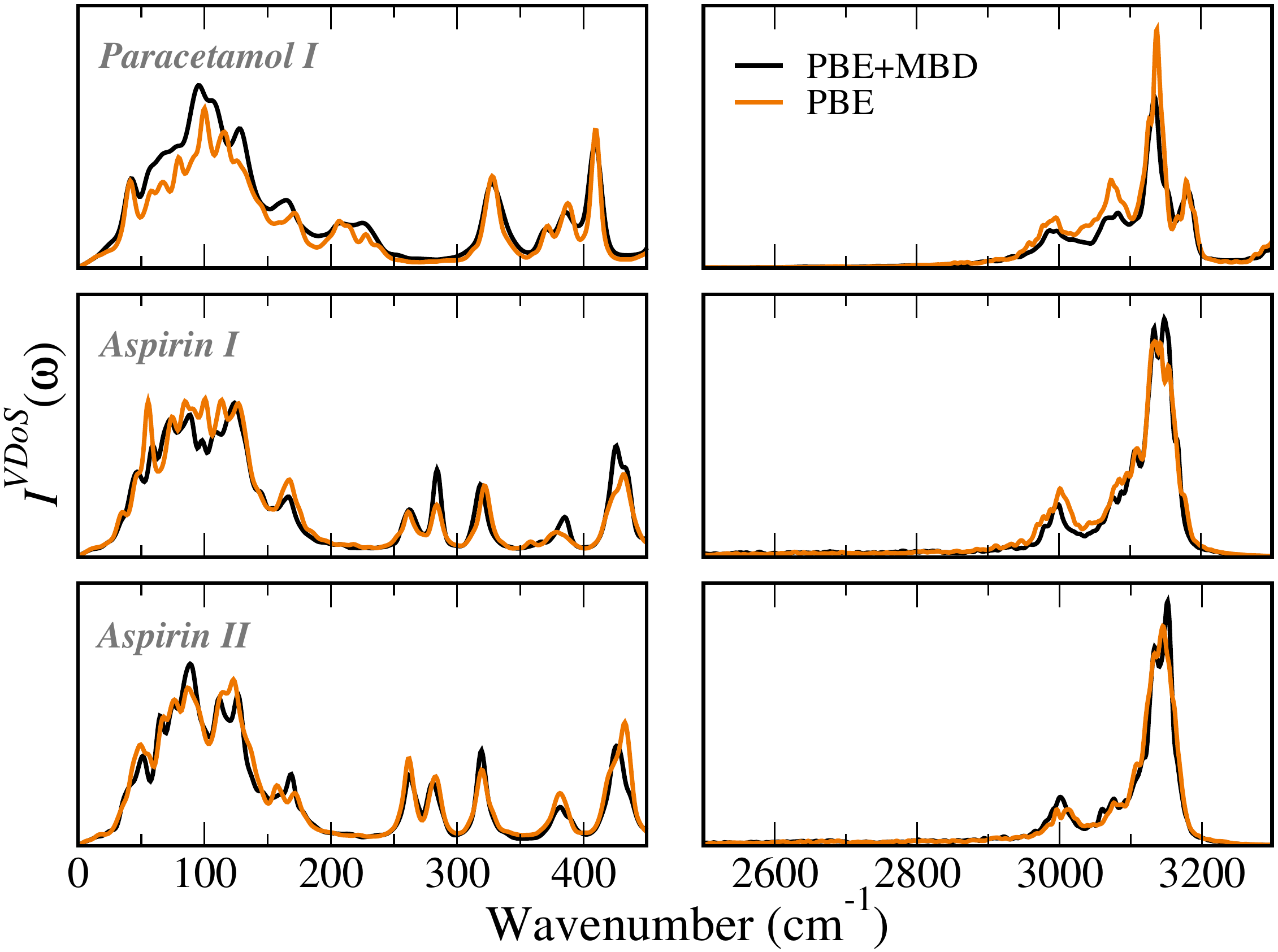}
  \caption{VDoS spectra of (from top to bottom) Paracetamol I, Aspirin I and Aspirin II obtained with the PBE (orange) and PBE+MBD (black) functionals.}
  \label{fig_vdos}
\end{figure}

We report the VDoS of Paracetamol I, Aspirin I and Aspirin II in Fig.~\ref{fig_vdos} with the PBE and PBE+MBD functionals (data for Paracetamol II and the PBE+MBD functional is shown in the SI\rev{, Fig. S5}). We note that in all cases, the VDoS is almost insensitive to the inclusion of many-body dispersion in the whole frequency range, \rev{an assertion that also holds for the harmonic VDoS when using the same lattice constants (see SI, Fig. S6)}.
It thus appears that anharmonic contributions play a more important role than the inclusion or not of vdW effects, \textit{if} the lattice parameters are kept constant (we stress that vdW are extremely important when it comes to relaxing unit cell parameters).

\begin{figure*}[ht]
  \includegraphics[width=2.0\columnwidth]{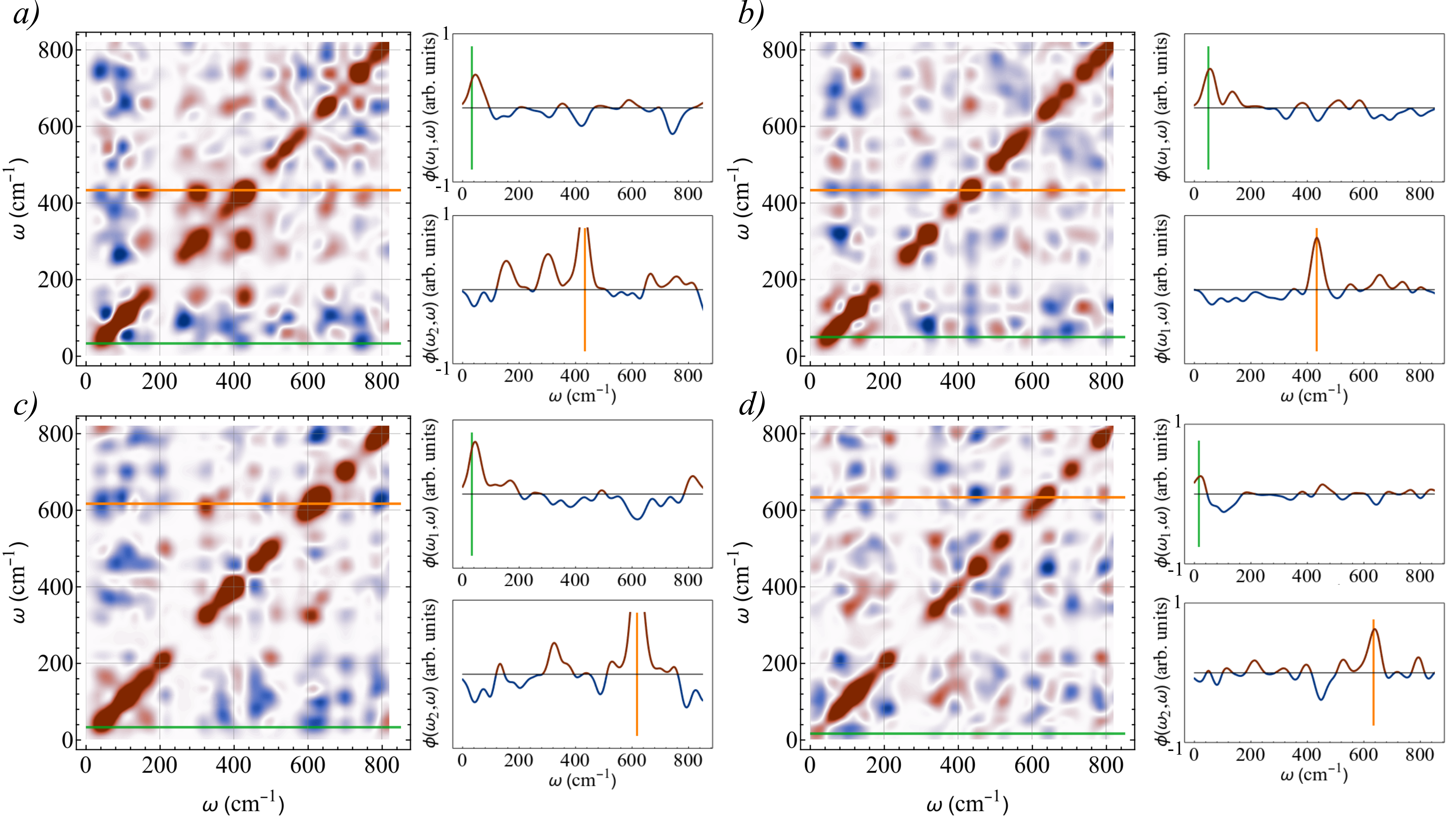}
  \caption{VDoS 2D correlation spectra of a) Aspirin I, b) Aspirin II, c) Paracetamol I and d) Paracetamol II. Blue (Red) indicates a negative (positive) correlation between modes, i.e., the intensities of a given pair vary in the opposite (same) direction. $\rev{T_{\text{win}}}=1$ ps.}
  \label{fig_2D_corr}
\end{figure*}

In order to characterize mode coupling, we calculate the 2D VDoS correlation spectra as explained in Section \ref{sec:2d}. Our results are shown in Fig.~\ref{fig_2D_corr} for all four polymorphs presented in this study. For each 2D-correlation plot, we make two cuts. The first cut corresponds to the lowest observed frequency of the system, while the second one is meant to highlight a difference in inter-mode couplings between polymorphs. Additionally, we focus on correlations within the 0-900cm$^{-1}$ region only. The following assignment of vibrational modes will be based on results from the harmonic approximation, albeit knowing that frequency shifts may be present. We choose each time the most probable eigenmode, i.e., the one corresponding to the closest harmonic eigenfrequency.

\rev{Overall, we observe that the vibrational modes showing stronger frequency shifts in Fig.~\ref{fig_lowrange_poly} are also the ones showing stronger correlations with other modes in Fig.~\ref{fig_2D_corr}. For example, for Aspirin I the mode at 125 cm$^{-1}$ which marks the edge of the intense low-frequency Raman signal is considerably shifted (towards experiment) in the anharmonic case, and it is seen to show a pronounced anti-corrrelations around 250 and 600 cmm$^{-1}$. Another example is the peak labeled $e$ of Paracetamol I in Fig.~\ref{fig_lowrange_poly}. It shows several pronounced anti-correlations around 366 cm$^{-1}$ (bending motion of the benzene ring), 465 cm$^{-1}$ (C-O bendings in the benzene plane and rocking of the benzene) and 630 cm$^{-1}$ (breathing mode of the benzene rings). In the following, we discuss other noteworthy aspects of these 2D correlations, that evidence differences between the sets of polymorphs.}

For Aspirin I, the lowest-observed frequency at around 35 cm$^{-1}$, corresponds to a ``sliding-motion" of the molecules with respect to one another. It couples in particular to a high-frequency mode at about 740 cm$^{-1}$ that mainly consists of a wagging motion of the benzene ring. 
The second vibration we focus on, at around 430 cm$^{-1}$, corresponds to collective motions involving in majority CO and CC bending motions; It couples positively most strongly to two other modes at ~155 and ~310 cm$^{-1}$, corresponding, respectively, to methyl group rotations (rocking) and to bendings between methyl groups and their radicals.

For Aspirin II, the 35 cm$^{-1}$ mode (analogue to that of Aspirin I) couples most strongly (and positively) to the mode at 130 cm$^{-1}$ which corresponds to collective partial methyl group rotations and bending motions between benzene rings and their radicals. 
The second cut at 430 cm$^{-1}$ (analogous to that of Aspirin I), shows only weak correlation to other modes, in stark contrast to Aspirin I.

For Paracetamol I, the lowest frequency vibration around 35 cm$^{-1}$ (involving small rotations of the individual molecular units as a whole) shows a pronounced coupling with another mode at 630 cm$^{-1}$, consisting for the most part of C-C bendings. 
Similarly, the second mode we pick at 630 cm$^{-1}$, shows in particular two couplings at 35 cm$^{-1}$ and 800 cm$^{-1}$, the latter consisting of N-H, C-H and O-H bending motions, without twisting the backbone structure.

For Paracetamol II one observes that in the low-frequency region, the lowest energy mode at about 16 cm$^{-1}$ (``sliding-motion" of the molecules with respect to one another) couples very little to higher-frequency vibrations.
Conversely, the intense mode at 630 cm$^{-1}$, which is extremely similar to that of Paracetamol I, has a strong negative coupling with a mode at ~450 cm$^{-1}$, composed mainly of C-O and C-N bendings.

All the vibrational modes mentioned above can be visualized in the SI, \rev{Figs. S10--S13}.
In general, there is no unique coupling between two specific vibrational modes, but rather a complex pattern of correlations between several of them in this low-frequency range composed of delocalized modes. This serves as a guide to understand which polymorphs and in which frequency regions one can expect more changes due to anharmonic effects and can thus serve as a diagnostic tool as to whether harmonic evaluations of free energy will be more or less accurate. We plan to further explore this aspect in the future. 

\section{Conclusions}

In this paper, we calculated harmonic and anharmonic Raman spectra of two polymorphs of Paracetamol and Aspirin, using a recent implementation of DFPT in the FHI-aims code and focusing especially in the low-frequency range, below 300 cm$^{-1}$.
We studied the impact of quasi-harmonic lattice expansion over harmonic Raman spectra, and concluded that while the middle- and high-frequency ranges are almost insensitive to moderate changes in lattice parameters, the low-frequency harmonic Raman spectra show important changes.
We also measured the influence of many-body dispersion corrections, both in harmonic and anharmonic Raman spectra and vibrational density of states at fixed experimental lattice parameters.
\rev{In the harmonic picture, the impact of MBD for both Raman and VDoS spectra is almost negligible. In the anharmonic picture at room temperature, the impact of adding MBD interactions is overall very tenuous.}
Dispersion interactions are, nevertheless, extremely important for determining the lattice parameters and the thermal lattice expansion in these crystals, as one would expect, given the nature of the intermolecular interactions. We compared anharmonic Raman spectra below 300 cm$^{-1}$ to experimental room-temperature spectra of all polymorphs, obtaining \rev{good} agreement. Finally we reported VDoS 2D-correlation spectra, and showed that different correlations exist between low- and higher-frequency vibrational modes, suggesting a high degree of anharmonicity for specific modes, but not for others. Interestingly, similar vibrational modes in different polymorphs show very different correlation patterns to other modes.

Overall, our results show that vibrational properties calculated from aiMD can accurately describe the low-frequency as well as the high-frequency vibrational region of molecular crystals, \rev{and reproduce the finer lineshape structure, provided that enough statistical sampling is performed.} \rev{The harmonic approximation does reproduce the main experimental peaks as well, even though several peaks are shifted with respect to anharmonic results, especially for Paracetamol. We observe that the most pronounced peak-shifts in the low-frequency range correlate with stronger off-diagonal correlation in our 2D-correlation spectra.} This means that temperature-dependent free energy calculations based on aiMD \cite{Mariana_PRL_2016} can indeed serve as benchmark values for such crystals, provided the cost of such simulations is affordable and a good estimation of the lattice constants at different temperatures is possible, either through simulations or experiment. 

The results also highlight once more the complexity of studying systems like molecular crystals, the structure and properties of which depend on a delicate interplay between several phenomena.
\rev{Cheaper methods like the harmonic approximation give very valuable insights into these structures, but are by essence bound to fail for anharmonic modes and high temperatures. In the present study, the most dramatic failure of the harmonic approximation was observed in the high-frequency OH-stretch mode of the Aspirin crystals.}
The calculation of 2D-correlation spectra could allow, in principle, to assess the validity of harmonic free-energy evaluations, given that weak intermode correlations suggest that anharmonic effects are less important. This aspect could be exploited in the future. 

For a more thorough understanding and assessment of polymorphic molecular crystals, the anharmonic route thus seems to be unavoidable if one can overcome the cost of such simulations in large scale studies. \rev{As an example, the calculation of the trajectories needed to obtain the anharmonic Raman spectrum of  Paracetamol I, required about 300000 core-hours (around 75000 for the dynamics, and 225000 for the computation of polarizabilities) in the COBRA supercomputer (processor type Intel Skylake 6148). This cost breakdown suggests that efficiently modelling the electric-field response properties could represent a larger gain in time than employing a good model to calculate the forces.}
Machine-learning approaches \rev{can be employed in this case, as will be demonstrated in an upcoming article~\cite{GPR_prep}. Such a reduction in computational cost will allow the study of a much wider range of systems.}

\section{Acknowledgments}

N.R. thanks Jonathan Burley for valuable comments about experimental Raman intensity corrections. M.R. thanks Michele Ceriotti for insightful comments at early stages of this manuscript. V.A. thanks funding from the WISE program of the Deutscher Akademiker Austauschdienst (DAAD). \rev{N.R. and M.R. thank Joscha Hekele for revising the code used to produce harmonic Raman spectra.}

\bibliography{main} 
\bibliographystyle{apsrev4-1}

\end{document}